\documentclass[10pt,preprint]{aastex}

\newcounter{thefigs}

\newcounter{thetabs}

\newcounter{address}

\def\simless{\mathbin{\lower 3pt\hbox
	{$\,\rlap{\raise 5pt\hbox{$\char'074$}}\mathchar"7218\,$}}} 
\def\simgreat{\mathbin{\lower 3pt\hbox
	{$\,\rlap{\raise 5pt\hbox{$\char'076$}}\mathchar"7218\,$}}} 

\slugcomment{Submitted to \aj}

\begin{document}
 

\title{An Efficient Algorithm for Positioning Tiles \\
in the Sloan Digital Sky Survey}


\author{
Michael R. Blanton\altaffilmark{\ref{Fermilab}},  
Robert H.~Lupton\altaffilmark{\ref{Princeton}},  
F.~Miller Maley\altaffilmark{\ref{PrincetonCS}},  
Neal Young\altaffilmark{\ref{Dartmouth}},  
Idit Zehavi\altaffilmark{\ref{Fermilab}},  
Jon Loveday\altaffilmark{\ref{Sussex}}
}
\setcounter{address}{1}
\altaffiltext{\theaddress}{
\stepcounter{address}
Fermi National Accelerator Laboratory, P.O. Box 500,
Batavia, IL 60510
\label{Fermilab}}
\altaffiltext{\theaddress}{
\stepcounter{address}
Princeton University Observatory, Princeton,
NJ 08544
\label{Princeton}}
\altaffiltext{\theaddress}{
\stepcounter{address}
Department of Computer Science, Princeton University,
Princeton, NJ 08544
\label{PrincetonCS}}
\altaffiltext{\theaddress}{
\stepcounter{address}
Department of Computer Science,
Dartmouth College,
6211 Sudikoff Laboratory,
Hanover, NH 03755-3510
\label{Dartmouth}}
\altaffiltext{\theaddress}{
\stepcounter{address}
Sussex Astronomy Centre,
University of Sussex,
Falmer, Brighton BN1 9QJ, UK
\label{Sussex}}


\begin{abstract}
The Sloan Digital Sky Survey (SDSS) will observe around $10^6$ spectra
from targets distributed over an area of about 10,000 square degrees,
using a multi-object fiber spectrograph which can simultaneously
observe 640 objects in a circular field-of-view (referred to as a
``tile'') 1.49$^\circ$ in radius. No two fibers can be placed closer
than $55''$ during the same observation; multiple targets closer than
this distance are said to ``collide.'' We present here a method of
allocating fibers to desired targets given a set of tile centers which
includes the effects of collisions and which is nearly optimally
efficient and uniform.  Because of large-scale structure in the galaxy
distribution (which form the bulk of the SDSS targets), a na{\"\i}ve
covering the sky with equally-spaced tiles does not yield uniform
sampling. Thus, we present a heuristic for perturbing the centers of
the tiles from the equally-spaced distribution which provides more
uniform completeness. For the SDSS sample, we can attain a sampling
rate of $>92\%$ for all targets, and $>99\%$ for the set of targets
which do not collide with each other, with an efficiency $>90\%$
(defined as the fraction of available fibers assigned to targets).
\end{abstract}

%
%

\section{Introduction}
\label{intro}

The Sloan Digital Sky Survey (SDSS; \citealt{york00a}) is producing a
deep imaging survey over about 10,000 square degrees, using a camera
with a large format CCD array on a dedicated telescope at Apache Point
Observatory in New Mexico (\citealt{gunn98a}). A sample of objects
selected from this imaging survey is being targeted for a
spectroscopic follow-up survey which is being conducted
concurrently. About 900,000 of these spectroscopic targets will be
galaxies (\citealt{strauss01a}), about 100,000 will be QSOs
(\citealt{richards01a}), and about 100,000 will be selected by color
to be intrinsically very red, luminous galaxies known as ``Luminous
Red Galaxies'' (\citealt{eisenstein01a}). In this paper, we will refer
to all of these objects generically as ``tiled targets,'' or often
simply ``targets.''

These targets are observed using two multi-object fiber spectrographs
on the same telescope (\citealt{uomoto01a}). Each spectroscopic fiber
plug plate, referred to as a ``tile,'' has a circular field-of-view
with a radius of 1.49$^\circ$ and can accommodate 640 fibers, 48 of
which are reserved for observations of blank sky and
spectrophotometric standards. Because of the finite size of the fiber
plugs, the minimium separation of fiber centers is $55''$. If, for
example, two objects are within $55''$ of each other, both of them can
be observed only if they lie in the overlap between two adjacent
tiles. Simulations and early observations both suggest that 10\% of
targets in the SDSS will be unobservable if they do not lie in
overlaps of tiles; about 30\% of the sky will be covered by such
overlaps.
The goal of the SDSS is to observe $99\%$ of the maximal set of
targets which has no such collisions (about 90\% of all targets). In
Section \ref{sdss} we give a more complete description of the details
of the SDSS.

Around 2,000 tiles will be necessary to provide fibers for all the
targets in the survey. Since each tile which must be observed
contributes to the cost of the survey (due both to the cost of
production of the plate and to the cost of observing time), we desire
to minimize the number of tiles necessary to observe all the desired
targets. In order to maximize efficiency (defined as the fraction of
available fibers assigned to tiled targets) when placing these tiles
and assigning targets to each tile, we need to address two
problems. First, we must be able to determine, given a set of tile
centers, how to optimally assign targets to each tile --- that is, how
to maximize the number of targets which have fibers assigned to
them. This problem is non-trivial because the circular tiles overlap.
Second, we must determine the most efficient placement of the tile
centers, which is non-trivial because the distribution of targets on
the sky is non-uniform, due to the well-known clustering of galaxies
on the sky. It turns out that the first problem can be solved in
polynomial time, even in the presence of fiber collisions (as long as
targets are distributed across the sky in a reasonable way). The
second problem belongs to a class of problems for which only
exponentially expensive methods for finding the exact solution are
known (that is, it is ``$NP$-complete''), but we use a heuristic
method developed by \citet{lupton98a} to find an approximate solution.

This paper discusses the strategy used by the SDSS to place its tiles
using these methods. It is designed to run on a patch of sky
consisting of a set of rectangles in a spherical coordinate system,
known in SDSS parlance as a ``chunk.'' Much of the strategy was
described by \cite{lupton98a}; this paper provides more astronomical
context and describes the method for resolving fiber
collisions. Section \ref{method} describes the method.  Section
\ref{tests} shows example results from actual SDSS data and from
simulations. Section \ref{sdss} describes some technical aspects of
the SDSS.  Section \ref{conclusion} summarizes our results.

\section{Tile Placement and Fiber Allocation}
\label{method}

Here we describe our method for placing each tile and allocating the
fibers of each tile to the targets. First, we discuss the allocation of
fibers given a set of tile centers. In the absence of fiber
collisions, this problem can be solved quickly and optimally, as shown
in Section \ref{nocollide}. A method which is nearly optimal in the
presence of fiber collisions is presented in Section
\ref{collide}. Second, in Section \ref{tileplace}, we discuss how to 
efficiently place the centers of the tiles.

\subsection{Target-to-Tile Assignment without Collisions}
\label{nocollide}

Given a distribution of targets on the sky and an {\it a priori} set
of tile centers, one can find the optimal solution to the problem of
allocating the targets to each tile, such that the maximum possible
number of targets are assigned fibers. With circular tiles, which
necessarily overlap, this is a somewhat non-trivial problem.

\begin{figure}
\epsscale{0.5}
\plotone{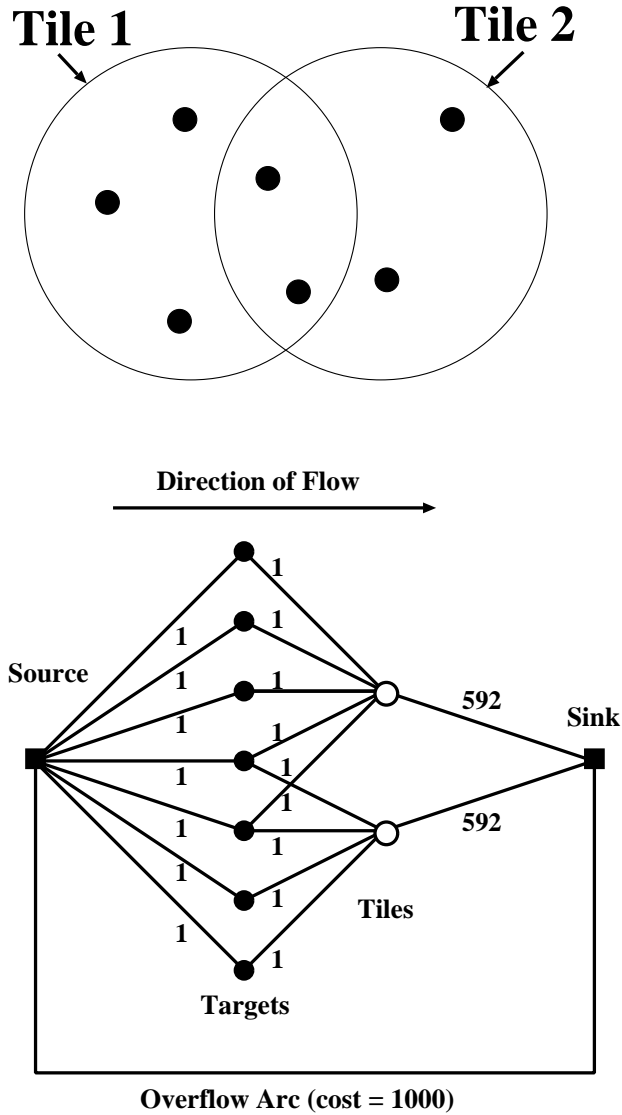}
\caption{\label{simple} Top half shows a schematic distribution of
targets (black dots) and the placement of two tiles used to observe
these targets. Bottom half shows the network flow which would be
constructed to optimally assign the targets to each tile. Each black
dot is a ``node;'' in analogy to the top half, the filled dots
represent targets and unfilled dots represent tiles.  Each line is an
``arc.'' The arcs are each labeled by a number which represents their
``capacity.'' Unless
otherwise marked, there is no cost associated with allowing targets to
flow down an arc. One should imagine that each target in the top panel
contributes to the flux of some fluid flowing from the source at
left. In this analogy, each arc is like a pipe which can accommodate
some maximum flow, and the nodes are locations where these pipes join,
and where the flow can be redirected. We want to direct the flow of
all the targets to the sink at right for the lowest possible
cost. Since the direct route (the ``overflow arc'') from the source to
the sink, which does not flow through any tile nodes and thus
corresponds to not observing a galaxy, has a substantial cost, the
minimum cost requirement effectively means maximizing the number of
targets which are assigned to tiles. }
\end{figure}

Figure \ref{simple} shows at the top a very simple example of a
distribution of targets and the positions of two tiles we want to use
to observe these targets. Given that for each tile there is a finite
number of available fibers, how do we decide which targets get
allocated to which tile?  As realized by \citet{lupton98a}, this
problem is equivalent to a network flow problem, which computer
scientists have been kind enough to solve for us already ({\it e.g.},
\citealt{goldberg97a}).

The basic idea is shown in the bottom half of Figure \ref{simple},
which shows the appropriate network for the situation in the top
half. Using this figure as reference, we here define some terms
which are standard in combinatorial literature and which will be
useful here:
\begin{itemize}
\item {\tt node}: The nodes are the solid dots in the figure; they 
provide either sources/sinks of objects for the flow or simply serve
as junctions for the flow. For example, in this context each target
and each tile corresponds to a node. 
\item {\tt arc}: The arcs are the lines connecting the nodes. They
show the paths along which objects can flow from node to node.  In
Figure \ref{simple}, it is understood that the flow along the arc
proceeds to the right. For example, the arcs traveling from target
nodes to tile nodes express which tiles each target may be assigned
to.
\item {\tt capacity}: The minimum and maximum capacity of each arc is
the minimum and maximum number of objects that can flow along it. For
example, because each tile can accommodate only 592 fibers, the
capacities of the arcs traveling from the tile nodes to the sink node
is 592.
\item {\tt cost}: The cost per object along each arc is exacted for
allowing objects to flow down a particular arc; the total cost is the
summed cost of all the arcs. In this paper, the network is designed
such that the minimum total cost solution is the desired solution.
\end{itemize}
Imagine that you have a flow of 7 objects which enters the network at
the source node at the left. The goal is for the entire flow to leave
the network at the sink node at the right for the lowest possible
cost. The objects must travel along the arcs, from node to node. Each
arc has a maximum capacity of objects which it can transport, as
labeled.  (One can also specify a {\it minimum} number, which will be
useful later). Each arc also has an associated cost, which is exacted
per object which is allowed to flow across that arc.  Arcs link the
source node to a set of nodes corresponding to the set of
targets. Each target node is linked by an arc to the node of each tile
it is covered by. Each tile node is linked to the sink node by an arc
whose capacity is equal to the number of fibers available on that
tile. None of these arcs has any associated cost. Finally, an
``overflow'' arc links the source node directly to the sink node, for
targets which cannot be assigned to tiles. The overflow arc has
effectively infinite capacity; however, a cost is assigned to objects
flowing on the overflow arc, guaranteeing that the algorithm fails to
assign targets to tiles only when it absolutely has to.  This network
thus expresses all the possible fiber allocations as well as the
constraints on the numbers of fibers in each tile.  Finding the
minimum cost solution (which can be done in polynomial time using the
method of \citealt{goldberg97a}) then maximizes the number of targets
which are actually assigned to tiles.

However, there are a couple of properties of the network flow
solutions which must be treated with caution. First, notice that in
this example there are only three types of target nodes: those only in
tile 1, those only in tile 2, and those in both.\footnote{The astute
reader will notice that there is therefore a more (computationally)
efficient way of setting up the network than shown in Figure
\ref{simple}, and indeed \citet{lupton98a} describe doing so. We
implement the more costly method in this situation because it is
simpler and we can afford it computationally.}  When the network flow
algorithm we use here chooses its solution, it does not guarantee that
it chooses targets within each of these types randomly. Thus, if the
order of the target nodes as sent to the network flow algorithm are
correlated with any target property (for example, position on the
sky), the distribution of that property in targets assigned to tiles
will differ from the distribution in all of the targets. Take for
example the situation that the targets are sorted by right
ascension. If the algorithm is unable to assign fibers to some of the
targets, it is likely that the unassigned targets will be nodes that
are close to each other in Figure \ref{simple}.  Therefore, they
will also be clumped in right ascension. This is unacceptable if we
desire a reasonable window function.  Thus, we randomize the order in
which nodes are assigned to targets by this algorithm; this prevents
any correlation between target properties and whether a target gets a
fiber.

Second, the particular method we use, provided by \citet{goldberg97a},
has the interesting property that when a certain number of targets
cannot be allocated fibers, the algorithm preferentially chooses to
exclude targets which are covered by more than one tile. This property
has no effect on the overall efficiency of the solution, but because
the method for fitting for tile positions presented in Section
\ref{tileplace} will tend to put overlaps of tiles in preferentially
overdense regions, it may introduce subtle correlations between the
sampling rate and the density field. This behavior is important if the
level of completeness in low is parts of the tiling region; however,
the uniformity of our completeness is high enough that this effect is
not important.

\subsection{Target-to-Tile Assignment with Collisions}
\label{collide}

As described above, there is a limit of $55''$ to how close two fibers
can be on the same tile. If there were no overlaps between tiles,
these collisions would make it impossible to observe $\sim 10\%$ of
the SDSS targets.  Because the tiles are circular, some fraction of
the sky will be covered with overlaps of tiles, allowing some of these
targets to be recovered.  In the presence of these collisions, the
best assignment of targets to the tiles must account for the presence
of collisions, and strive to resolve as many as possible of these
collisions which are in overlaps of tiles.  We approach this problem
in two steps, for reasons described below. First, we apply the network
flow algorithm of Section \ref{nocollide} to the set of ``decollided''
targets --- the largest possible subset of the targets which do not
collide with each other. Second, we use the remaining fibers and a
second network flow solution to optimally resolve collisions in
overlap regions.

\subsubsection{Network Flow for Decollided Objects}
\label{decnetflow}

The effect of fiber collisions is one which any analysis of the SDSS
data is going to face. The fact that most ($70\%$) of the sky in the
survey will be only covered by a single tile means that a certain
number of objects will be missed for this reason. Thus, the best that
one can hope for in terms of sampling is that all unobserved targets
have a close neighbor which {\it was} observed; the redshift of
the observed target would thus give us some prior information on the
redshift of the unobserved target. Furthermore, exactly where and how
many targets one can recover in tile overlaps depends strongly on the
locations of the tiles, and thus on the target density field. To
evaluate the effect of this dependence on large-scale structure
statistics, one would have to run the algorithm described here on a
large number of mock catalogs. For these two reasons, we must take
care that we identify a set of targets which we could observe no
matter where the tiles are, and obtain as complete a sample as
possible of {\it these} targets.

\begin{figure}
\epsscale{0.8}
\plotone{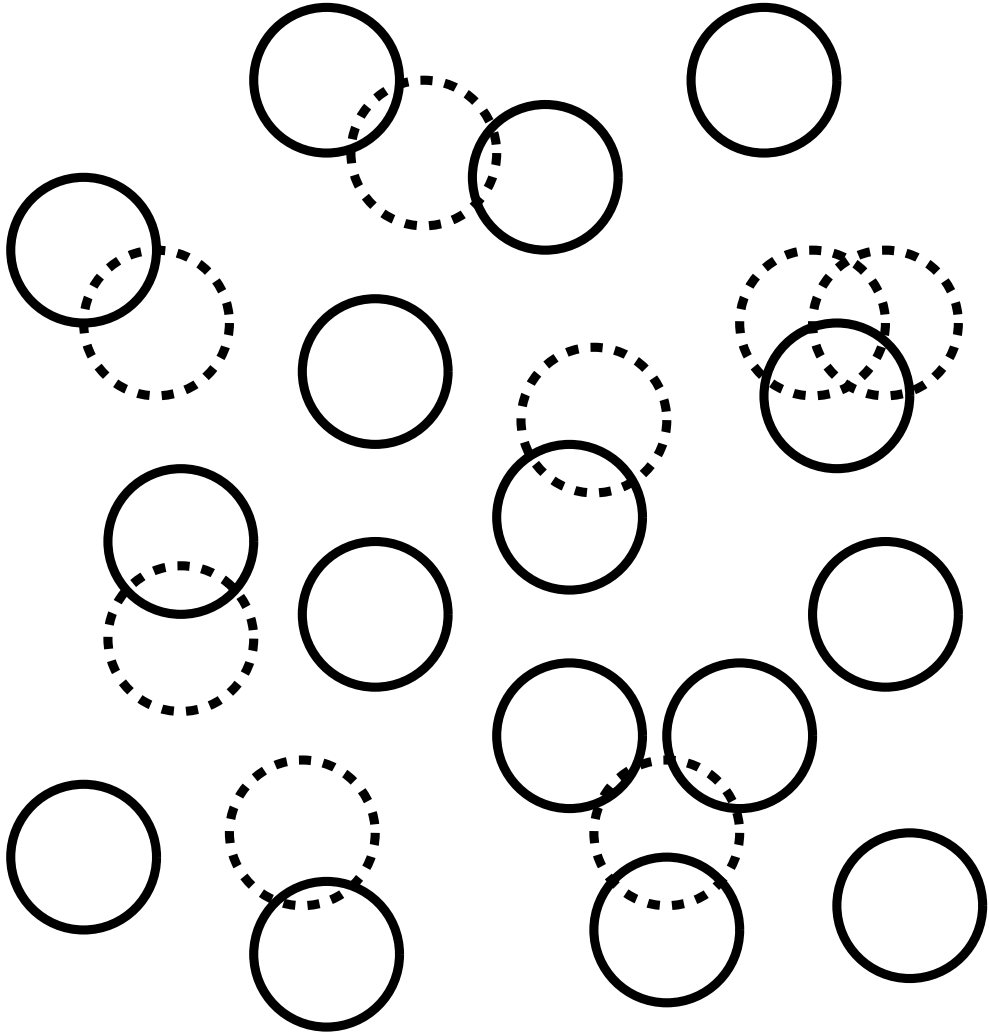}
\caption{\label{maximal} Dramatization of the definition of the
``decollided'' set of galaxies. Each circle (both solid and dashed) is
centered on the location of a target and has a diameter equal to the
fiber collision limit. Thus, intersecting circles represent targets
which ``collide.'' The solid circles represent the largest subset of
galaxies which can be chosen which do not ``collide'' with each
other. We refer to these galaxies as a ``decollided'' set of
galaxies. Note that there is usually no unique decollided set, because
(for example) in a binary collision we are always free to choose
either galaxy to be decollided.}
\end{figure}

To identify this sample, we define the maximal subset of the targets
which are all greater than $55''$ from each other, which we refer to
as the ``decollided'' set. To clarify what we mean by this maximal
set, consider Figure \ref{maximal}. Each circle represents a target;
the circle diameter is $55''$, meaning that overlapping circles are
targets which collide. The set of solid circles is the ``decollided''
set. Thus, in the triple collision at the top, it is best to keep the
outside two rather than the middle one. To find this decollided set of
targets, we run a friends-of-friends grouping algorithm on the targets
with a $55''$ linking length. The resulting groups are almost always
of sufficiently low multiplicity that we can simply check all
possibilities to find the best possible selection of targets which
eliminates fiber collisions. We pick at random one of the set of
equivalent ``best'' selections; for example, if two objects collide,
this algorithm simply picks one at random to be ``decollided.''

This determination is complicated slightly by the fact that some
targets are assigned higher priority than others. For example, as
explained in Section \ref{tiledtargets}, QSOs are given higher priority than
galaxies by the SDSS target selection algorithms. What we mean here by
``priority'' is that a higher priority target is guaranteed never to
be eliminated from the sample due to a collision with a lower priority
object. Thus, our true criterion for determining whether one set of
assignments of fibers to targets in a group is more favorable than
another is that a greater number of the highest priority objects are
assigned fibers. In the case of a tie in the highest priority objects,
the next highest priority objects are considered, and so on.

Once we have identified our set of decollided objects, we use the
network flow solution to find the best possible assignment of fibers
to that set of objects.


\subsubsection{Network Flow for Collisions}

After allocating fibers to the set of decollided targets, there will
usually be unallocated fibers, which we want to use to resolve
fiber collisions in the overlaps. We can again express the problem of
how best to perform the collision resolution as a network, although
the problem is a bit more complicated in this case. In the case of
binaries and triples, we design a network flow problem such that the
network flow solution chooses the tile assignments optimally. In the
case of higher multiplicity groups, our simple method for binaries and
triples does not work and we instead resolve the fiber collisions in a
random fashion; however, fewer than 1\% of targets are in such groups,
and the difference between the optimal choice of assignments and the
random choices made for these groups is only a small fraction of that.

The design of the second network flow is similar to the first, with
source and sink nodes (connected by the overflow arc) and a layer of
tile nodes.  However, instead of a layer of {\it target} nodes, we
have a layer of nodes corresponding to each {\it group} (as defined by
the aforementioned friends-of-friends algorithm) which has at least
one member in an overlap of tiles. We thus ignore groups covered only
by one tile, since we have already done as well as possible for those
targets. Note we include single-member groups in this process, which
allows targets in overlaps which were previously guaranteed a fiber on
one plate to be shuffled to another plate, if it proves desirable to
do so.

First, we need to set the properties of each arc connecting the source
node to each group node. For each group, we find the maximum number of
targets which could be observed, $c_{\mathrm{max}}$, taking advantage of
the overlapping tiles, but regardless of the number of available
fibers in each tile. We find $c_{\mathrm{max}}$ by simply trying all
possible target-to-tile configurations for that group, and picking the
best solution (again accounting for the relative priorities of the
objects).  A constraint on the best solution is that some subset of
the targets in each group will have been allocated fibers as
decollided targets in the first network flow. Any ``best'' solution
must guarantee that these targets will be assigned fibers again in the
second network flow. In addition, these required targets clearly set a
minimum number of targets to observe in each group,
$c_{\mathrm{min}}$. Each source-to-group arc will have its maximum and
minimum capacity set according to these bounds.

Second, we need to set the properties of each arc connecting a group
node to a tile node.  In the case that $c_{\mathrm{max}}\le 3$, we
determine the maximum number of targets which can be assigned to each
tile, $c_{\mathrm{max},i}$, given all the equivalent ``best'' sets of
target-to-tile assignments. The minima $c_{\mathrm{min},i}$ are set to
the minimum number of arcs in each tile, given all legal
target-to-tile assignments.  The group-to-tile arcs are then assigned
these maximum and minimum capacities.  
Under these conditions, in almost every case, any solution the network
flow finds will be achievable, in the sense that it can be implemented
without the occurrence of fiber collisions. See below for a discussion
of the exceptions.

In the case that $c_{\mathrm{max}}>3$, the same prescription does not
provide the assurance that the network flow will return a viable set
of tile assignments, and instead we pick a particular ``best'' set of
target-to-tile assignments for each such group in order to guarantee
feasibility. In this case, the code chooses at random a particular
realization of the ``best'' resolution of the fiber collisions, (one
must also be careful, in the $c_{\mathrm{max}}>3$ case, to guarantee
fibers to all of the decollided fibers which were picked in the first
network flow solution; this task is complicated but tractable).

\begin{figure}
\epsscale{0.5}
\plotone{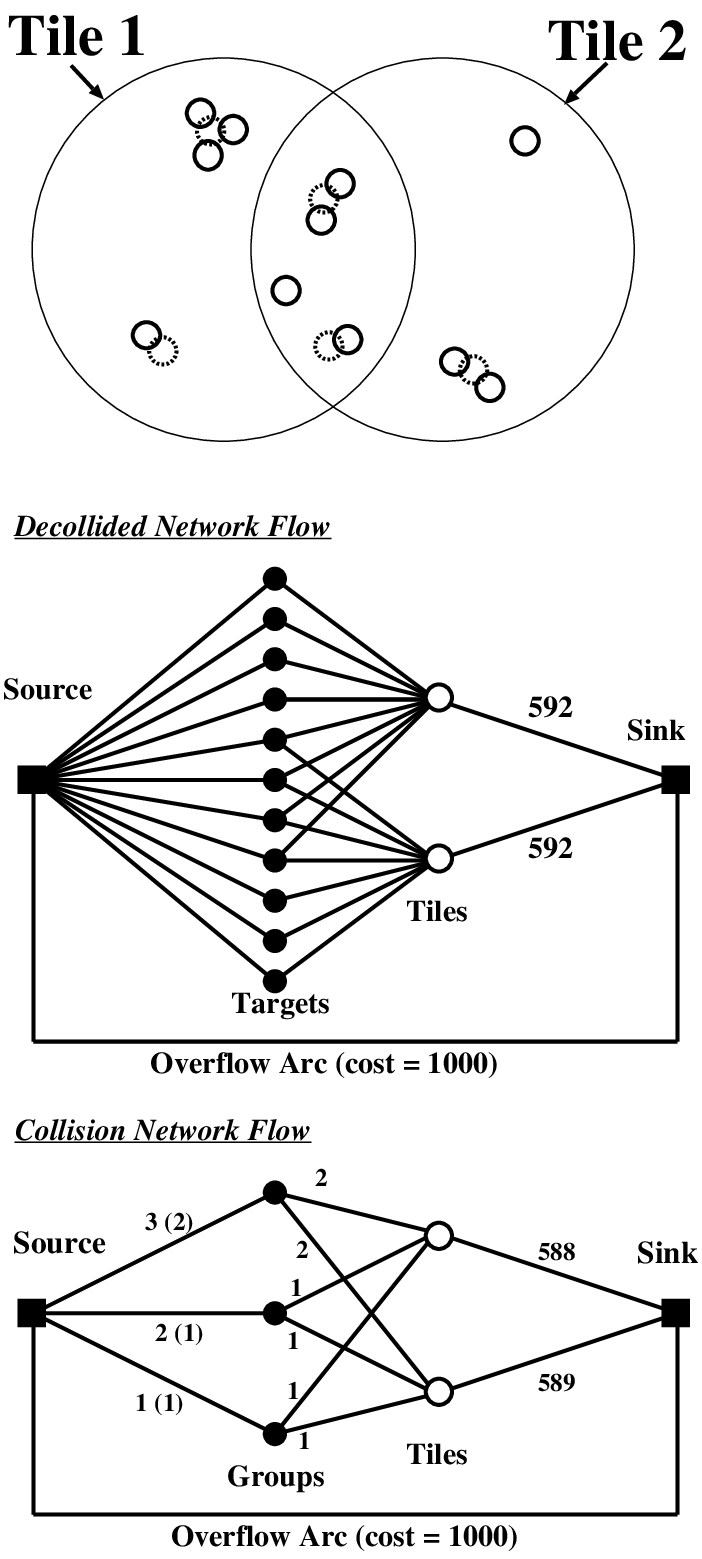}
\caption{\label{collisions} Top panel shows a schematic distribution
of targets and tiles, as in Figure \ref{simple}; in the manner of
Figure \ref{maximal}, the decollided galaxies are solid circles and
the others are dashed circles. Middle panel shows the decollided
network flow (Section \ref{decnetflow}) used to find the optimal
solution for decollided objects; this network flow has the same form
as that in Figure \ref{simple} (here unmarked arcs have a capacity of
unity). Bottom panel shows the network flow used to resolve collisions
in overlaps of tiles. In this case, the set of target nodes has been
replaced by nodes corresponding to each group with one or more members
in an overlap of tiles. For the case shown here, there are three such
groups. The arcs to and from each group have minimum and maximum
capacities set as described in the text. If omitted, the maximum
capacity is unity. The minimum capacity for each arc is put in
parentheses after the maximum; if omitted, the minimum capacity is
zero. }
\end{figure}

For the sake of concreteness, consider Figure \ref{collisions}, which
shows a possible tile-target configuration and the networks which
would be constructed to solve it. Again, the solid circles indicate
the ``decollided'' set of targets, which has 11 members, and for which
the decollided network flow is run (unmarked arcs have a capacity of
unity). Assuming all the decollided targets are obtained, we set up
the network flow for the groups in the overlap as shown at the
bottom. Each source-to-group arc is marked by its maximum capacity
$c_{\mathrm{max}}$ followed by its minimum capacity $c_{\mathrm{min}}$
in parentheses. As explained above, these minima are set by the fact
that some of the targets are guaranteed spots because they were
previously assigned tiles in the decollided solution. The
group-to-tile capacities are set to the maximum possible on any given
tile.  Again, setting things up this way allows the network flow
solution to optimally allocate the overlap fibers (at least for
triples and binaries) while still guaranteeing that a solution is
possible and that fibers are assigned to all the decollided targets
which had been previously selected.  

As mentioned above, there are cases for which these rules return
unfeasible answers. Under the conditions of the SDSS, these
cases are extraordinarily rare; we mention them because the same may
not be true for every application. There are essentially two classes
of failures. First, occasionally it happens that because part of a
group is in an overlap and part is not, the ``best'' solution {\it
requires} that more fibers be assigned to a tile than were assigned in
the decollided solution. If, in conjunction with this occurrence, the
tile in question is in a particularly dense region, it may already
require all its fibers to cover the decollided targets. Thus,
applying the rules above creates a second network flow with no
possible solution. In such cases, the code reverts to a ``fail-safe''
mode which only allows solutions to the group-to-tile problem which put
the same number of decollided targets onto each tile as were assigned
in the first network flow. 

Second, and again very rarely, it occasionally happens that while the
second network flow successfully returns a choice of fiber
assignments, this choice makes it impossible to assign fibers to all
the decollided targets which were guaranteed fibers in the first
network flow, not because of a lack of fibers but because of
geometrical considerations. Again, the problem is associated with
groups which straddle tile boundaries.  In this case, the problem
occurs when some targets in a group are in an overlap of three or
more tiles, and others to be in an overlap of a lesser number of
tiles. In the code, we simply warn the user that some decollided
targets have been lost. On the basis of simulations, we expect ten to
twenty of the million SDSS targets to be lost due to this effect.

\subsection{Tile Placement}
\label{tileplace}

Once one understands how to assign fibers given a set of tile centers,
one can address the problem of how best to place those tile centers.
One can show that to solve this problem optimally is $NP$-hard ({\it
e.g.}~\citealt{megiddo84a}), but \citet{lupton98a} have developed a
heuristic method which works well for the sorts of distributions of
targets we deal with here. This method first distributes tiles
uniformly across the sky and then uses a cost-minimization scheme to
perturb the tiles to a more efficient solution.

\subsubsection{Initial Conditions}

We need to choose some initial, nearly uniform covering of the region
to be tiled, before perturbing it to improve the efficiency. We use
two techniques. First, for sufficiently large chunks of sky, we draw
the uniform tiling from an approximately uniform covering of the
sphere provided by \citet{hardin00a}. These coverings are provided for
discrete numbers of tiles; the choice appropriate for the SDSS target
density (about 120 per square degree) is 7682 tiles over the whole
sky. We throw away tiles whose centers are not in the chunk of sky of
interest to us. Second, for smaller chunks of sky (which a small chunk
of the uniform spherical covering is less likely to cover in a
reasonable way) we simply lay down a rectangle of tiles, with the
centers of the tiles in each row offset in order to provide a complete
covering.

\subsubsection{Perturbing the Tiles}
\label{perturb}

The method is essentially iterative. One starts with a uniform
covering of tiles over the region in question, as described in the
previous subsection. Then, one allocates targets to the tiles, but
instead of limiting a target to the tiles within a tile radius, one
allows a target to be assigned to further tiles, but with a certain
cost which increases with distance (remember that the network flow
accommodates the assignment of costs to arcs).  For group-to-tile
nodes in the second network flow solution, one defines the cost
according the position of the group center. One uses exactly the same
fiber allocation procedure as above.

In practice, we do not allow fibers to be assigned to any tile, but
only those within $2.5$ times the tile radius. We assign a cost of the
following form:
\begin{equation}
c = \begin{array}{cc}
0 & r < R_{\mathrm{tile}}\cr
A \left[ \left(\frac{r}{R_{\mathrm{tile}}}\right)^\alpha - 1 \right]
& r > R_{\mathrm{tile}}
\end{array},
\end{equation}
where $r$ is the distance of the fiber from the center of the tile,
$R_{\mathrm{tile}}$ is the radius of the tile, and $\alpha$ is the
logarithmic slope of the cost function. $A$ is a scale factor, set so
that at $r=2.5 R_{\mathrm{tile}}$ the cost is equal to the cost of not
assigning the fiber at all.

What this does is to give each tile some information about the
distribution of targets outside of it. Then, once one has assigned a
set of targets to each tile, one changes each tile position to that
which minimizes the cost of its set of targets. To perform this
minimization, we use Powell's direction set method, as described by
\citet{press92a}. Then, with the new positons, one reruns the fiber
allocation, perturbs the tiles again, and so on. As \citet{lupton98a}
point out, this method is guaranteed to converge to a minimum (though
not necessarily a global minimum), because the total cost must
decrease at each step.

The parameter $\alpha$ sets the slope of the cost function; the most
advantageous value of $\alpha$ depends in detail on the density and
distribution of the targets.  We generally set $0.5 < \alpha <
2$. High values in this range encourage tiles to take large excursions
from their initial positions, since the slope of the cost function
becomes higher at larger radii. Under these conditions, tiles are
influenced by distant targets that they may never cover; however, this
behavior can be desirable for large chunks of sky for which the best
solution may require large numbers of tiles to shift in unison. Low
values in this range are more conservative in the sense that tiles are
encouraged to travel less far from their initial positions, since the
slope of the cost function decreases with radius. This behavior is
usually desirable for small chunks of sky, for which many tiles are
sitting near an edge and large changes of position will usually
uncover sky. Perhaps a more general approach is to allow $\alpha$ to
be variable in some way throughout the minimization.


Depending on the overall survey goals, one can choose to which set of
targets these costs apply. For the SDSS, we are most interested in
maximizing the fraction of decollided targets which are observed. For
this reason, we assign cost {\it only} to the decollided targets,
effectively ignoring the other objects when fitting for tile
positions. In fact, during the iteration we do not even perform the
second network flow. 

It is possible to assign fibers to a slightly larger fraction (by
about $1\%$) of all targets if all targets are included in the cost
minimization. However, for the SDSS this improvement would come at the
cost of large numbers of gaps opening up between tiles, because the
number of tiles necessary to observe all the targets is uncomfortably
close to the number of tiles necessary to simply cover the available
sky. This effect highlights an important facet of the tiling problem:
inefficiency arises because tiles which are in underdense regions
cannot always be moved towards dense regions without leaving parts of
the sky completely uncovered. A much higher target (and thus tile)
density would mitigate this difficulty.

In practice, we also need to determine the appropriate number of tiles
to use. Thus, using a standard binary search, we repeatedly run the
cost-minimization to find the minimum number of tiles necessary to
satisfy the SDSS requirements, namely that we assign fibers to $>99\%$
of the decollided targets.

\section{Testing the Method}
\label{tests}

In order to test how well this algorithm works, we apply it both to
simulated and real data. First, we test the algorithm on a large
solid angle sample drawn from an $N$-body simulation. Second, we show
results based on actual tiling solutions for a small chunk of sky in
the SDSS commissioning data.

\subsection{Simulation Tests} 

\begin{figure}
\epsscale{1.0}
\plotone{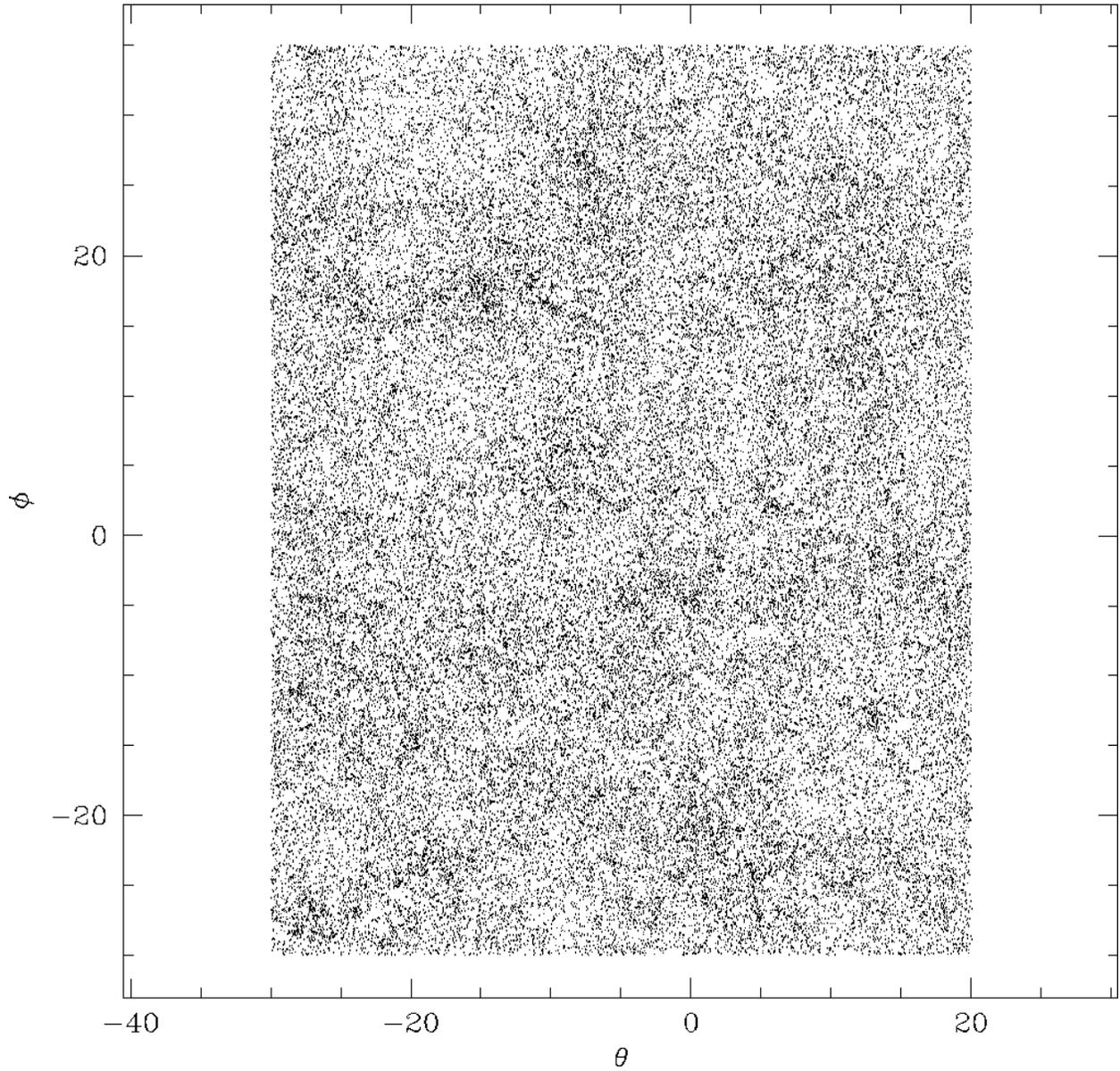}
\caption{\label{simgalaxies} Distribution of targets on the sky, using
galaxies drawn from a simulation by \citet{cole98a}, plus 20 targets
per square degree added randomly to represent LRG and QSO targets. We
have subsampled the targets by a factor of five for this plot.}
\end{figure}

For this exercise, we use the simulations of \citet{cole98a}, which
are collisionless $N$-body simulations of the growth of structure in a
COBE-normalized Cold Dark Matter (CDM) model with $\Omega_m = 0.3$,
$\Omega_\Lambda = 0.7$, and $\sigma_8 = 1.05$. In this simulation,
dark matter particles are chosen randomly to represent galaxies, and
are assigned luminosities based on an assumed luminosity function. The
location of an observer is chosen, a flux limit is assumed, and the
galaxies in the simulation are ``observed.'' The resulting
distribution of galaxies has about the same redshift distribution as
do galaxies in the actual SDSS survey, with a median $z\sim 0.1$. This
procedure results in a surface density (about 90 per square degree)
and an angular clustering of galaxies on the sky approximately the
same as the SDSS. In order to simulate the quasar and LRG samples, we
distribute an extra 20 targets per square degree randomly on the sky;
although in three dimensions both populations are highly clustered,
their large distance and sparse sampling make the approximation that
they are randomly distributed in angle not bad for our purposes.  We
extract a section of the simulation about 3075 square degrees in solid
angle (a rectangle in spherical coordinates spanning the latitude
range $-30^\circ < \theta < 35^\circ$ and the longitude range
$-30^\circ < \phi < 20^\circ$) and consisting of 336,392 objects.  The
distribution of galaxies in this range
is given in Figure \ref{simgalaxies}. This angular region is probably
larger than any that will be available during the course of the SDSS.

\begin{figure}
\epsscale{1.0}
\plotone{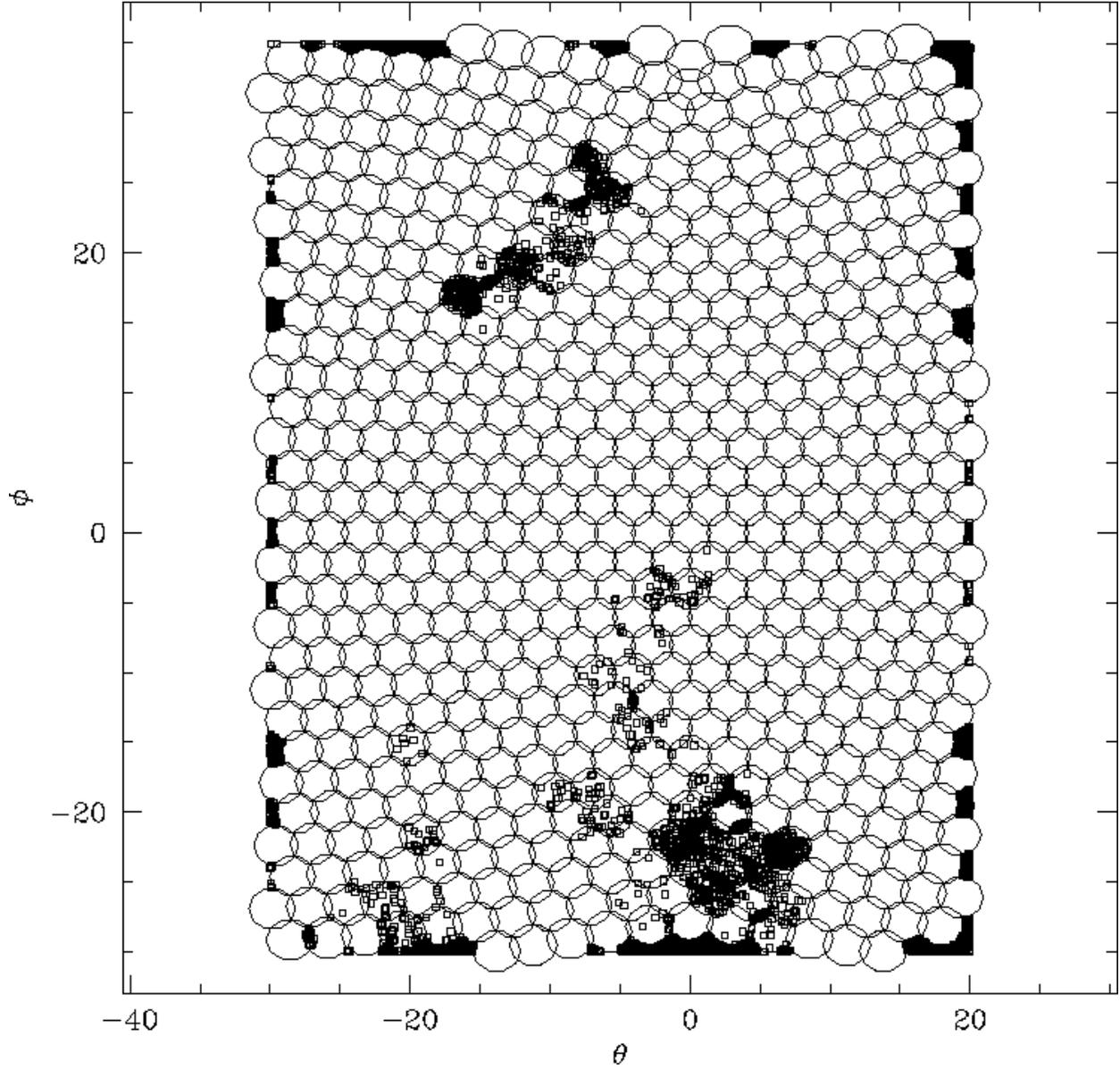}
\caption{\label{simuniform_dec} Results of distributing tiles
uniformly across the targeted region. The boundaries of the tiles are
shown. Missing decollided galaxies are shown as squares. While overall
the completeness is high, note that in the densest regions, many
decollided objects are missing, with the completeness becoming as low
as 10\% in the most incomplete regions. }
\end{figure}

As initial conditions for this large ``chunk,'' we extract a portion
of a nearly uniform covering of the sphere given by
\citet{hardin00a}. We exclude any tiles whose centers are outside the
official boundaries of the chunk. This procedure will leave missing
targets near the edges; these targets can be recovered when the
adjacent region of sky is tiled. In any case, any gaps which are left
when the survey is completed can be accounted for in the window
function, to the extent that those gaps are uncorrelated with the
underlying density of galaxies. For our first test, we do not perturb
the positions of the tiles at all, and assign the fibers to the
uniformly distributed tiles. The results are shown in Figure
\ref{simuniform_dec}; here we show the tiles as circles and the
missing decollided galaxies as squares.  Decollided galaxies which
were assigned fibers and all collided galaxies are omitted from the
plot. The statistics associated with this solution are given in the
first column of Table
\ref{simtable}. It is clear that although the overall completeness is
high ($\sim 98.3\%$ of decollided objects are assigned) the small
amount of incompleteness is concentrated in a few, dense regions of
sky. The patch of incompleteness near the bottom center is about
85--90\% complete in the decollided objects on average; the most
incomplete sections of that are only 10---30\% complete. Clearly it is
unsatisfactory to have such a high rate of incompleteness concentrated
in unusually dense regions of sky, even if the overall completeness of
the survey is high. Such a strong correlation of the sampling fraction
with the galaxy density field poses difficulty estimating large-scale
structure statistics.

%
%

\begin{figure}
\plotone{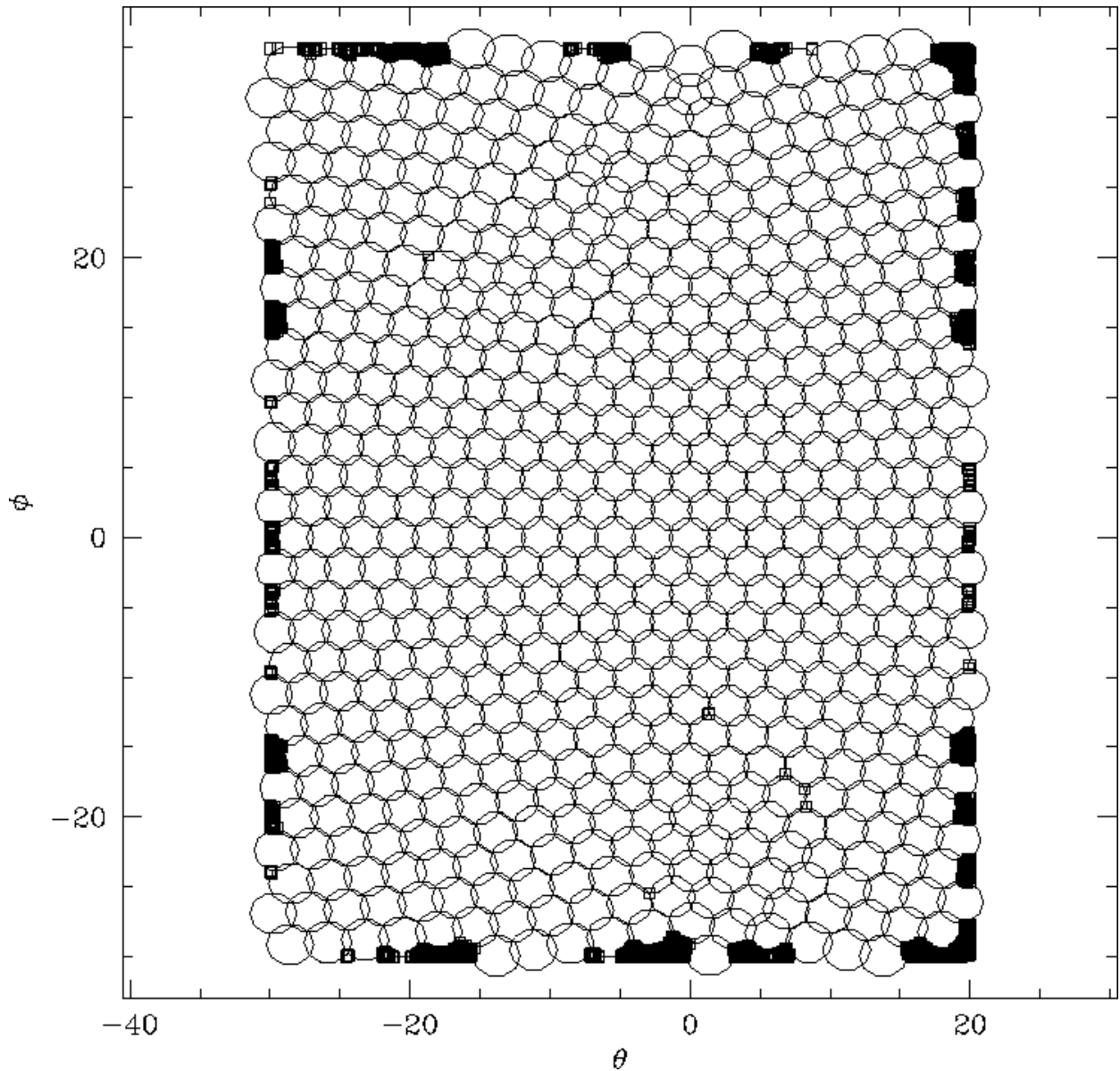}
\caption{\label{simperturbed_dec} Same as Figure \ref{simuniform_dec},
now with the results of perturbing the positions of the tiles using
the cost minimization heuristic described in Section
\ref{perturb}. While the tiles move very little from their uniform
initial distribution, the completeness has improved and has become far
more uniform.  Only a few objects are missing in the interior of the
sample.  This improvement occurs because in the densest regions tiles
are pushed together, and thus overlap more.  }
\end{figure}

Let us therefore perturb the tiles in an effort to increase the
completeness and its uniformity. The result of applying the method
described in Section \ref{tileplace} is shown in Figure
\ref{simperturbed_dec}; the resulting tiling statistics are given in
the second column of Table \ref{simtable}. (Note that one extra tile
was added in the process; this has a negligible effect on the
statistics in Table
\ref{simtable}). Now there are only a handful of objects missing in
the interior of the sample. All of the missing objects are
concentrated at the edges. Thus, while the overall completeness is
increased a bit (to $\sim 99.0\%$ of decollided objects), the real
improvement is in the uniformity of the sample.

\begin{figure}
\plotone{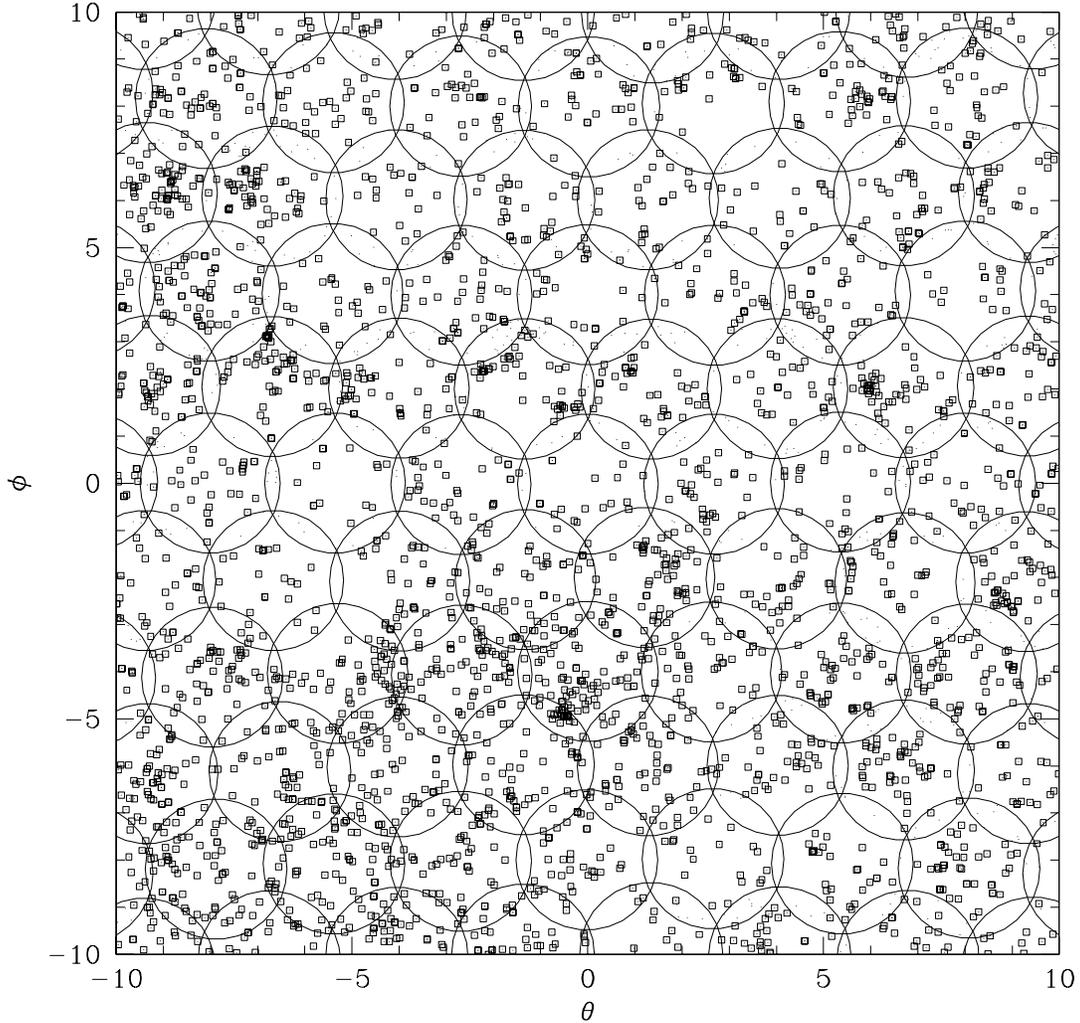}
\caption{\label{simperturbed_col} Here we have zoomed in on a section
of Figure \ref{simperturbed_dec}. In this figure the points are the
collided object ({\it i.e.} those objects that are not in the
decollided set). Open squares are placed over those collided objects
which do {\it not} receive fibers; that is, they show objects in fiber
collisions which did not get resolved.  Obviously all objects bumped
by collisions are missed in regions covered by a single tile. When
extra fibers are available, as happens near the top of the figure,
almost all of the fiber collisions in the overlaps of tiles are
resolved. Of course, when all the fibers are used on decollided
objects in the first network flow, as happens near the bottom of the
figure, none are left to resolve collisions in overlaps.}
\end{figure}

To show how the resolution of collisions works, we show as points in
Figure \ref{simperturbed_col} the collided galaxies (those which are
not in the decollided set).  We have zoomed into a section of the
interior to make it easier to distinguish the points.  These points
represent the set of objects which would be eliminated due to fiber
collisions if there was no overlap between tiles.  Open squares are
placed over those which did not receive fibers.  It is clear
that in the underdense regions most of the collisions in overlaps of
tiles are actually resolved. In the overdense regions, however, almost
all of the fibers are used to observe decollided targets, and few are
left over to resolve collisions.

Overall, in these simulations, $\sim 92.5\%$ of the available targets
were assigned fibers; most of the missing ones are due to fiber
collisions which were not able to be resolved. The efficiency of the
tiling solution, quantified as the percentage of fibers which are used
on tiled objects, is $\sim 91.2\%$.

\subsection{Tests with SDSS Data} 


We here show tiling results using this method from SDSS commissioning
data. In this phase of the survey, there was not enough imaging data
yet to define a chunk as large as in the simulations of the previous
subsection, so we had to settle for a much smaller chunk of sky. This
chunk (known as ``Chunk 7'') of sky is 5 degrees wide and about 12
degrees long, and contains 6629 objects. It is the first chunk of SDSS
data on which this version of the code was used.

\begin{figure}
\plotone{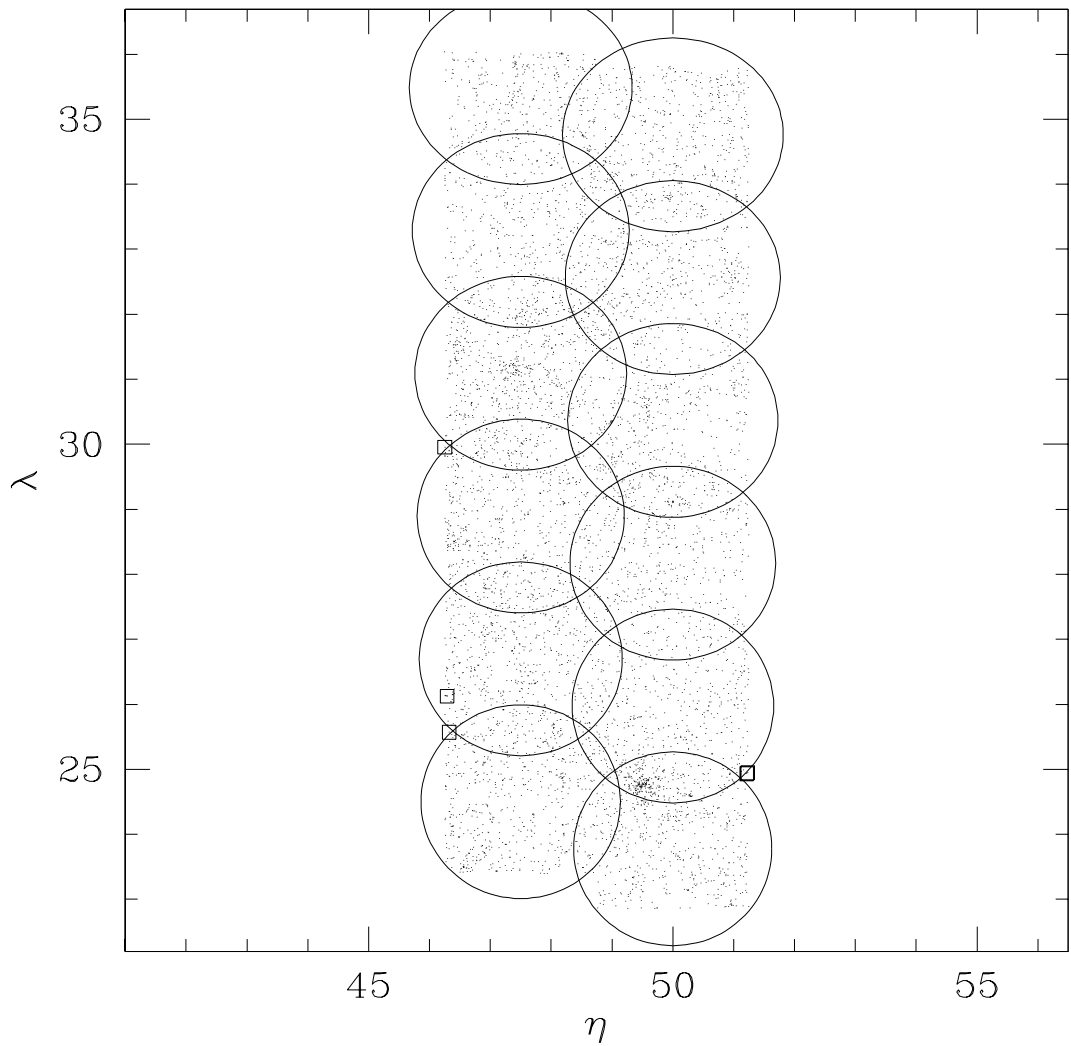}
\caption{\label{obsgalaxies} The distribution of targets in Chunk 7 of
the SDSS, displayed in ``survey coordinates.'' The positions of the
tiles are shown as well (they are nearly in the uniform positions in
which they were placed initially). The open squares show the five
decollided objects which were not assigned fibers. }
\end{figure}

The initial conditions were set up simply as a rectangular
distribution of tiles. The tile positions were perturbed in order to
maximize the number of decollided galaxies assigned to fibers.
However, in this case the tiles move very little --- the uniform
initial conditions turn out to be close to a minimum in our cost
function. The statistics of the solution are listed again in Table
\ref{simtable}; note that the efficiency is a bit low, mainly because
this chunk is small. The positions of the targets and tiles are given
in Figure \ref{obsgalaxies}.

An obvious criticism of the tiling of this chunk is that we should
only tile the center of the chunk, such that our tiles never cover sky
which has not yet been imaged. Then we would wait until later to
observe edges of the chunk. In terms of the total number of tiles
drilled, such an approach would be more efficient. However, doing this
would leave the telescope idle when it could be taking spectra, so it
is worth drilling a few more tiles than necessary in order to
optimally use the available time. In any case, the fibers left
unassigned to any main survey targets are assigned to other targets,
mainly stars, FIRST (\citealt{becker95a}) sources, and ROSAT
(\citealt{voges99a}) sources, so the unassigned fibers are by no means
wasted.

\section{Technical Details for SDSS Data}
\label{sdss}

There are a few technical details which may be useful to mention in
the context of SDSS data, since understanding these issues is crucial
to understanding the window function when calculating large-scale
structure statistics with the survey. First, we will describe which
targets within the SDSS are ``tiled'' in the manner described here,
and how such targets are prioritized. Second, we will discuss the
method used by SDSS to deal with the fact that the imaging and
spectroscopy are performed within the same five-year time
period. Third, we will describe the tiling outputs which the SDSS
tracks as the survey progresses.  Throughout, we refer to the code
which implements the algorithm described above as {\tt tiling}.

The information described in this section (along with the
spectroscopic results) is necessary but not quite sufficient to
calculate large-scale structure statistics for the survey. First, at
later stages in the processing, fibers can be lost due to collisions
with guide fibers, as well as with the center of the tile, where a
post prevents any fiber from being placed within $100''$ (in later
versions, we will adjust the algorithm described here to attempt to
avoid placing tile centers so close to targets).  Second, some fields
within each chunk are excluded for reasons such as bad
seeing. Finally, bright stars make it impossible to observe galaxies
in a certain fraction of the sky, in a way which varies with Galactic
latitude. These masks need to be determined to study clustering on the
largest scales in the survey.

\subsection{Targets Which are ``Tiled''}
\label{tiledtargets}

Only some of the spectroscopic target types identified by the target
selection algorithms in the SDSS are ``tiled.'' These types (and their
designations in the primary and secondary target bitmasks, as
described in \citealt{stoughton01a}) are listed in Table
\ref{tiled}. They consist of most types of QSOs, main sample galaxies,
LRGs, hot standard stars, and brown dwarfs. These are the types of
targets for which tiling is run and for which we are attempting to
create a well-defined sample. Once the code has guaranteed fibers to
all possible ``tiled targets,'' remaining fibers are assigned to other
target types by a separate code.

All of these target types are treated equivalently, except that they
assigned different ``priorities,'' designated by an integer.  As
described above, the tiling code uses them to help decide fiber
collisions. The sense is that a higher priority object will never lose
a fiber in favor of a lower priority object.  The priorities are
assigned in a somewhat complicated way for reasons immaterial to
tiling, but the essense is the following: the highest priority objects
are brown dwarfs and hot standards, next come QSOs, and the lowest
priority objects are galaxies and LRGs. QSOs have higher priority than
galaxies because galaxies are higher density and have stronger angular
clustering. Thus, allowing galaxies to bump QSOs would allow
variations in galaxy density to imprint themselves into variations in
the density of QSOs assigned to fibers, which we would like to avoid.
For similar reasons, brown dwarfs and hot standard stars (which have
extremely low densities on the sky) are given highest priority.

Each tile, as stated above, is 1.49$^\circ$ degrees in radius, and has
the capacity to handle 592 tiled targets. No two such targets may be
closer than 55$''$ on the same tile.

\subsection{Definition of a Tiling Chunk}

The {\it modus operandi} of the SDSS makes it impossible to tile the
entire 10,000 square degrees simultaneously, because we want to be
able to take spectroscopy during non-pristine nights, based on the
imaging which has been performed up to that point. In practice,
periodically a ``chunk'' of data is processed, calibrated, has targets
selected, and is passed to the tiling code. During the first year of
the SDSS, about one chunk per month has been created; as more and more
imaging is taken and more tiles are created, we hope to decrease the
frequency with which we need to make chunks, and to increase their size.

The first chunk which is ``supported'' by the SDSS is denoted Chunk
4. The first chunk for which the version of tiling described here was
run is Chunk 7. Chunks earlier than Chunk 7 used a different (less
efficient) method of handling fiber collisions. The earlier version
also had a bug which artificially created gaps in the distribution of
the fibers. The locations of the known gaps are given in
\citet{stoughton01a} for Chunk 4, since it is part of the SDSS Early
Data Release.

A chunk is defined as a set of rectangles on the sky (defined in
survey coordinates; \citealt{stoughton01a}) on the sky.  All of these
rectangle are designed to cover only sky which has been imaged and
processed. Most of each chunk consists of targets which have not been
included in any previous chunk. However, if an earlier chunk was
adjacent, targets may have been missed near its edges because they
were not covered by tiles, so the areas near the edges of adjacent
chunks are also included. Thus, in general, chunks overlap.

\subsection{Tiling Outputs}

Once a chunk is tiled, the position of each tile is stored. The tiles
are assigned a global index for the survey known as a {\tt
tileId}. For each target, the {\tt tileId} to which it is assigned is
stored (or $-1$ if no fiber is assigned). In addition, the $55''$
group to which it belonged (indexed from 0 for each chunk
independently) is also stored as {\tt collisionGroup}. Finally, a {\tt
mask} parameter is created, whose three lowest bits are
(respectively): {\tt ASSIGNED}, {\tt DECOLLIDED}, and {\tt
COVERED}. {\tt ASSIGNED} means that a fiber was actually assigned to
the target. {\tt DECOLLIDED} means that the target was designated a
decollided target. {\tt COVERED} means that the target was in an area
observable by some tile. Unfortunately, these parameters are not
included in the SDSS Early Data Release.

\section{Summary}
\label{conclusion}

This paper describes a method for positioning tiles and assigning
fibers to targets which is being used for the SDSS. The method assigns
fibers in a near-optimal manner, which is possible to do in polynomial
time given the sorts of target distributions found in the SDSS. We
note that if the typical nearest-neighbor distance of targets is of
order the fiber collision length, the groups found in the
friends-of-friends algorithm become very large, and the solution is
only possible in exponential time. The positioning of tiles is an
$NP$-complete problem (\citealt{megiddo84a}); we use the heuristic
devised by \citet{lupton98a} to find an approximate
solution. Importantly, we define a set of decollided targets for which
we can achieve nearly complete sampling; this fact will make the
survey easier to mimic when analyzing simulations. We have tested this
method both on simulations and on SDSS commissioning data. Finally, we
have described some of the technical details of the SDSS itself.

Variations of this method may be useful for future surveys consisting
of overlapping spectroscopic fields of view. The main lesson learned
in developing this method is that inefficiencies arise primarily due
to the need to completely cover the given area. To take the most
perverse possible case, if 592 objects were spread across the entire
sky, 592 tiles would be necessary. One would rather have those 592
targets within the area of a single tile, which could assign fibers to
all of them. Thus, to minimize the number of tiles drilled, one needs
a high enough target density that the number of tiles necessary to
observe the targets easily covers the survey area. This allows the
tiles more freedom to move to where they are most needed without
uncovering areas of sky in underdense regions; it also provides more
overlaps, and thus more ability to resolve fiber collisions. Since the
resulting tiling would be nearly 100\% efficient even for small
chunks, there would be no loss of efficiency due to the piecemeal
nature of the chunks. Because of its scientific goals, spectroscopic
instrumentation, and its budget, the SDSS is not in this optimal
regime. A large increase in target density (factor of two) would be
desirable from the point of view solely of tiling efficiency; however,
the survey goals and technical considerations make such a change
impossible. Naturally, we feel that the loss of efficiency is not
devastating, because the unused fibers are used to observe other
interesting targets, but we mention it here as an issue that future
surveys may wish to consider.

\acknowledgments

We would like to thank Daniel Eisenstein, Gillian Knapp, Don
Schneider, Ravi Sheth, Michael Strauss, and Daniel van den Berk for
advice and comments. MB is supported by the DOE and NASA grant NAG
5-7092 at Fermilab, and is grateful for the hospitality of the
Department of Physics and Astronomy at the State University of New
York at Stony Brook, who kindly provided computing facilities on his
frequent visits there.

The Sloan Digital Sky Survey (SDSS) is a joint project of The
University of Chicago, Fermilab, the Institute for Advanced Study, the
Japan Participation Group, The Johns Hopkins University, the
Max-Planck-Institute for Astronomy (MPIA), the Max-Planck-Institute
for Astrophysics (MPA), New Mexico State University, Princeton
University, the United States Naval Observatory, and the University of
Washington. Apache Point Observatory, site of the SDSS telescopes, is
operated by the Astrophysical Research Consortium (ARC).

Funding for the project has been provided by the Alfred P.~Sloan
Foundation, the SDSS member institutions, the National Aeronautics and
Space Administration, the National Science Foundation, the
U.S. Department of Energy, the Japanese Monbukagakusho, and the Max
Planck Society. The SDSS Web site is {\tt http://www.sdss.org/}.

\clearpage
\begin{deluxetable}{cccc}
\tablecolumns{4}
\tablecaption{\label{simtable} Tiling Results}
\tablehead{ & Simulation (Uniform) & Simulation (Perturbed) & SDSS Chunk 7
(Perturbed) }
\tablenotetext{a}{Fraction of targets which received fibers}
\tablenotetext{b}{Fraction of targets classified as decollided }
\tablenotetext{c}{Fraction of decollided targets which received
fibers}
\tablenotetext{d}{Fraction of collided targets in overlaps of tiles
which received fibers}
\tablenotetext{e}{Fraction of fibers assigned to targets}
\startdata
$N_{\mathrm{plates}}$ & 575 & 576 & 12\cr
$f_{\mathrm{tiled}}$\tablenotemark{a} & 0.918 & 0.924 & 0.933\cr
$f_{\mathrm{dec}}$\tablenotemark{b} & 0.919 & 0.919 & 0.902 \cr
$f_{\mathrm{tiled,dec}}$\tablenotemark{c} & 0.983 & 0.990 & 0.999 \cr
$f_{\mathrm{overlap}}$\tablenotemark{d} & 0.593 & 0.607 & 0.837 \cr
Efficiency\tablenotemark{e} & 0.907 & 0.912 & 0.870 \cr
\enddata
\end{deluxetable}

\begin{deluxetable}{lll}
\tablecaption{Target Selection Flags For Tiled Targets\label{tiled}}
\tablewidth{0pt}
\tablehead{
  \colhead{Name}        &  
  \colhead{Hex Bit}     &
  \colhead{Description}
}
\startdata
\\
\multicolumn{3}{l}{Primary Targets} \\
\\
  {\tt TARGET\_QSO\_HIZ}             & 0x1          &    
High-redshift QSO\\
  {\tt TARGET\_QSO\_CAP}             & 0x2          &    
QSO at high Galactic latitude\\
  {\tt TARGET\_QSO\_SKIRT}           & 0x4          &    
QSO at low Galactic latitude\\
  {\tt TARGET\_QSO\_FIRST\_CAP}      & 0x8          &    
``Stellar'' FIRST source at high Galactic latitude\\
  {\tt TARGET\_QSO\_FIRST\_SKIRT}    & 0x10         &    
``Stellar'' FIRST source at low Galactic latitude\\
  {\tt TARGET\_GALAXY\_RED}          & 0x20         &    LRG\\
  {\tt TARGET\_GALAXY}               & 0x40         &    Main sample galaxy\\
  {\tt TARGET\_GALAXY\_BIG}          & 0x80         &    Low surface brightness galaxy\\
  {\tt TARGET\_GALAXY\_BRIGHT\_CORE} & 0x100        &    
Low surface brightness galaxy with bright fiber magnitude\\
  {\tt TARGET\_STAR\_BROWN\_DWARF}   & 0x8000       &    Brown dwarf\\
\\
\multicolumn{3}{l}{Secondary Targets} \\
\\
  {\tt TARGET\_HOT\_STD}             & 0x200        &    Hot subdwarf standard star\\
\enddata
\end{deluxetable}

\clearpage
\clearpage

\setcounter{thefigs}{0}

\newpage
\stepcounter{thefigs}

\newpage
\stepcounter{thefigs}


\newpage
\stepcounter{thefigs}

\newpage
\stepcounter{thefigs}

\newpage
\stepcounter{thefigs}

\newpage
\stepcounter{thefigs}

\newpage
\stepcounter{thefigs}

\newpage
\stepcounter{thefigs}

\end{document}